\documentclass[sigconf]{acmart}

\pdfoutput=1 %

\usepackage[utf8]{inputenc}
\usepackage[most]{tcolorbox}

\usepackage{booktabs}
\usepackage{multirow}

\usepackage{graphicx}
\usepackage{setspace}
\usepackage{amsmath}
\usepackage{amsfonts} %
\usepackage{calligra} %
\usepackage{mathtools}

\usepackage{amsthm} %
\usepackage{tabularx}
\usepackage{rotating}

\usepackage{comment}
\usepackage{listings}
\usepackage{enumitem}

\usetikzlibrary{patterns}

\usepackage{blindtext}

\usepackage{acronym}

\newcommand{\code}[1]{{\small{\texttt{#1}}}}

\usepackage{float}
\usepackage{textcomp}

\usepackage{algorithm}
\usepackage{algorithmic}

\usepackage{url}

\usepackage[all=normal,paragraphs=tight, floats=tight,indent=tight]{savetrees}
\usepackage{framed}
\usepackage{soul}
\usepackage{csquotes}
\usepackage{multirow}
\usepackage{array}
\newcolumntype{L}[1]{>{\raggedright\let\newline\\\arraybackslash\hspace{0pt}}m{#1}}
\newcolumntype{C}[1]{>{\centering\let\newline\\\arraybackslash\hspace{0pt}}m{#1}}
\newcolumntype{R}[1]{>{\raggedleft\let\newline\\\arraybackslash\hspace{0pt}}m{#1}}
\newcolumntype{H}{>{\collectcell\lstinline}l<{\endcollectcell}}
\usepackage{url}
\MakeOuterQuote{"}
\usepackage[caption=false,font=footnotesize,labelformat=simple]{subfig} %

\definecolor{mygreen}{rgb}{0,0.6,0}
\definecolor{mygray}{rgb}{0.5,0.5,0.5}
\definecolor{mymauve}{rgb}{0.58,0,0.82}

\acrodef{CPS}{Cyber-Physical System}
\acrodef{IoT}{Internet of Things}
\acrodef{HDL}{Hardware Description Language}
\acrodef{CAD}{Computer-Aided Design}
\acrodef{EDA}{Electronic Design Automation}
\acrodef{HPC}{High-Performance Computing}
\acrodef{DL}{Deep Learning}
\acrodef{ML}{Machine Learning}
\acrodef{NLP}{Natural Language Processing}
\acrodef{IC}{Integrated Circuit}
\newcommand{\ignore}[1]{{}}

\newcommand{\squishlist}{
	\begin{list}{$\bullet$}
		{ \setlength{\itemsep}{0pt}
			\setlength{\parsep}{1pt}
			\setlength{\topsep}{1pt}
			\setlength{\partopsep}{0pt}
			\setlength{\leftmargin}{0.9em}
			\setlength{\labelwidth}{1.5em}
			\setlength{\labelsep}{0.4em} } }
	\newcommand{\squishend}{
	\end{list}  }

\definecolor{graphFirst}{RGB}{2,136,209} %
\definecolor{graphSecond}{RGB}{211,47,47} %
\definecolor{graphThird}{RGB}{245,124,0} %
\definecolor{graphFourth}{RGB}{56,142,60} %
\definecolor{graphFifth}{RGB}{81,45,168} %
\definecolor{graphSixth}{RGB}{69,90,100} %
\definecolor{graphSeventh}{RGB}{251,192,45} %
\definecolor{backgroundSecond}{RGB}{239,154,154} %
\definecolor{backgroundThird}{RGB}{255,204,128} %
\definecolor{backgroundFourth}{RGB}{165,214,167} %
\definecolor{backgroundFifth}{RGB}{179,157,219} %
\definecolor{backgroundSixth}{RGB}{176,190,197} %
\definecolor{backgroundSeventh}{RGB}{255,245,157} %

\copyrightyear{xx}
\acmYear{xx}
\setcopyright{none}\acmConference[xx 'xx]{xx}
\acmBooktitle{xx}
\acmPrice{xx.00}
\acmDOI{xx}
\acmISBN{xx}


\begin{document}

\title{A survey of Digital Manufacturing\\ Hardware and Software Trojans}

\author{Prithwish Basu Roy*, Mudit Bhargava, Chia-Yun Chang,\\ Ellen Hui, Nikhil Gupta, Ramesh Karri, Hammond Pearce}
\email{*pb2718@nyu.edu}
\affiliation{NYU Tandon School of Engineering}

\begin{abstract}
Digital Manufacturing (DM) refers to the on-going adoption of smarter, more agile manufacturing processes and cyber-physical systems. 
This includes modern techniques and technologies such as Additive Manufacturing (AM)/3D printing, as well as the Industrial Internet of Things (IIoT) and the broader trend toward Industry 4.0.
However, this adoption is not without risks: with a growing complexity and connectivity, so too grows the cyber-physical attack surface.
Here, malicious actors might seek to steal sensitive information or sabotage products or production lines, causing financial and reputational loss.
Of particular concern are where such malicious attacks may enter the complex supply chains of DM systems as Trojans---malicious modifications that may trigger their payloads at later times or stages of the product lifecycle.

In this work, we thus present a comprehensive overview of the threats posed by Trojans in Digital Manufacturing. We cover both hardware and software Trojans which may exist in products or their production and supply lines.
From this, we produce a novel taxonomy for classifying and analyzing these threats, and elaborate on how different side channels (e.g. visual, thermal, acoustic, power, and magnetic) may be used to either enhance the impact of a given Trojan or utilized as part of a defensive strategy.
Other defenses are also presented---including hardware, web-, and software-related.
To conclude, we discuss seven different case studies and elaborate how they fit into our taxonomy.
Overall, this paper presents a detailed survey of the Trojan landscape for Digital Manufacturing: threats, defenses, and the importance of implementing secure practices.

\end{abstract}

\maketitle

\thispagestyle{plain}
\pagestyle{plain}

\section{Introduction}
\label{sec:intro}

\subsection{Digital Manufacturing (DM) Overview}
Current industry practices promote the adoption of Digital Manufacturing (DM) techniques for more flexible production lines and faster time-to-market for products. 
These include cyber-physical systems (CPS) consisting of networks of computer-controlled manufacturing systems, which may be modern additive manufacturing (AM) processes (also referred to as 3D printing) or may be more traditional manufacturing techniques and systems newly modernized.
In general, DM systems make up the modern so-called `Industry 4.0'--- intelligent, fully integrated manufacturing environments containing both physical production line components alongside cyber/digital control~\cite{moller2016digital}.

Whereas traditional manufacturing sees specialized tooling produced for specific products and highly-coupled relationships between product designer and product manufacturer, DM instead allows for Manufacturing-as-a-Service (MaaS) where product designers and manufacturers may have a more flexible relationship. 
The ultimate goal is for digital objects such as design instructions, process recipes, and measurements may be transmitted over computer networks to end-users or a MaaS provider and produced on-demand~\cite{iquebal_towards_2018}.
This is particularly enabled by Additive Manufacturing (AM), where comparatively cheaper and general-purpose machines are beginning to manufacture deployment-quality parts and is increasingly being used in domains such as aerospace~\cite{Liao2006, Kobryn2001, geaviation}, automotive ~\cite{nyamekye2023impact} and biotechnology~\cite{heartvalves}. 
Overall, the AM market has increased exponentially from \$1.1B in 2009 to \$4.3B in 2016 and \$9.2B in 2019 at an annual growth rate of 26\%, with the market size being expected to double again within the next four years \cite{ref6_angrish2019fabsearch}.
The increasing demand for more complex and intricate designs at lower costs and faster rates than traditional manufacturing methods can allow making DM a rapidly growing field. Overall, the DM `Smart Manufacturing' domain is growing annually at 13.4\%, and after being globally valued at 254.24 billion USD in 2022, is set to exceed \$730B by 2030~\cite{noauthor_smart}.

\subsection{Threats to DM}
However, as the usage of DM systems increases, so does the potential for security threats. 
In general, attackers of DM will seek to either (1) gain sensitive knowledge about an attacked target, or (2) inflict damage (financial, physical, or reputational). 
The threat landscape is varied, and many cyber-based attacks may be defended against with traditional infosec `best-practices' (e.g. \cite{berkely_information_security_office_top_2023}), for instance, protecting against phishing with staff training plans, password hacking by 2-factor authentication, keeping software up to date, utilizing Virtual Private Networks (VPNs), firewalls, and anti-virus or anti-malware protections.

Other classes of attack are more difficult to defend against. Of particular concern are where threats may enter target products or production lines directly, perhaps via supply chain attacks, where they may remain dormant and difficult to detect until a trigger activates them---i.e., Trojans.
The introduction of hardware or software Trojans into DM systems has the potential for wide-ranging consequences, from the theft of information to the sabotage of products or entire production lines.
For example, recent works have shown that successfully inserted Trojans may impact the quality and strength of end-products~\cite{moore_implications_2017,moore2016vulnerability,pearce2022flaw3d}.
Other examples show that software Trojans may exfiltrate sensitive information~\cite{yampolskiy2021did}. 
Chettri et al.~\cite{chhetri2019tool}, show that digital AM systems are vulnerable to Trojans introduced through the software used to control the printer. These Trojans can be used to alter the printer's function or steal sensitive information, possibly by increasing the amount of information leaked through side-channels.

This potential for hardware and software Trojans in DM is particularly worrying in industries where the security of the products being manufactured is critical, such as in the automotive and aerospace domains~\cite{paritala_digital_2017}, pharmaceutical industry~\cite{drugs} and medical device industry~\cite{heartvalves, bones}. 
Preventing and detecting any malicious Trojans from insertion in these domains is paramount.

\subsection{Motivation and Contributions}
Realizing the threat, a number of surveys addressing cyber-attacks on DM and its subset AM have been published in the last five years (e.g. \cite{gupta2020additive,mahesh2020survey,challengesin3d,yampolskiy2016using,tuptuk_security_2018}). 
These surveys have introduced their own taxonomy and have suggested possible attacks and defenses that are primarily hypothetical, covering a broad scope of possible physical, cyber-physical, and purely cyber-based threats in nature. 
In our work we focus exclusively on classifying the risks, identifying the target goals, and presenting potential defensive strategies against hardware and software Trojans targeting Digital Manufacturing.
We limit ourselves to Trojans primarily due to their potentially long-lasting and severe consequences, as well as the difficulty to prevent them via traditional information security practices.
Due to their closely related nature, we also include details on side-channels and how Trojans might emphasize these (for information leakage) or be detected via side-channel analysis.
As it focuses on hardware and software Trojans, we consider attacks and defenses relevant to the Computer Aided Design (CAD) and Computer Aided Manufacturing (CAM) phases of a product's life cycle. In our case studies, we discuss in depth some of these attacks and existing ways of defending DM against them.

The contributions of this work are as follows:
\begin{itemize}[leftmargin=0.3cm]
    
    \item We discuss the cybersecurity risks of Digital Manufacturing in the presence of Hardware and Software Trojans.
    
    \item We present in detail the demonstrated detection and countermeasure techniques against Trojans which may target DM.
    
    \item We define a Threat Model and taxonomy to categorize Trojan threat goals, threat vectors, and countermeasures.

    \item We discuss this taxonomy in the context of seven different Trojan-based case studies, demonstrating its classification capabilities.
\end{itemize}

The rest of this paper is structured as follows: In Section~\ref{sec:preliminaries}, we provide an overview of Trojans and their intersection with the security risks associated with DM. 
In Section~\ref{sec:threatmodel}, we formalize a novel taxonomy for classifying and analyzing the various threats to AM systems. 
Section~\ref{sec:detection} then discusses existing detection techniques for these threats. 
Section~\ref{sec:case-studies} presents five case studies to illustrate the practical applications of the concepts discussed in the previous sections. 
Finally, we conclude the paper in Section~\ref{sec:conclusions} with a summary of the key findings and suggestions for future research.

\section{Background}
\label{sec:preliminaries}
\subsection{Hardware and Software Trojans}
\begin{figure}[t]
\centering
\centerline{\includegraphics[scale=0.7]{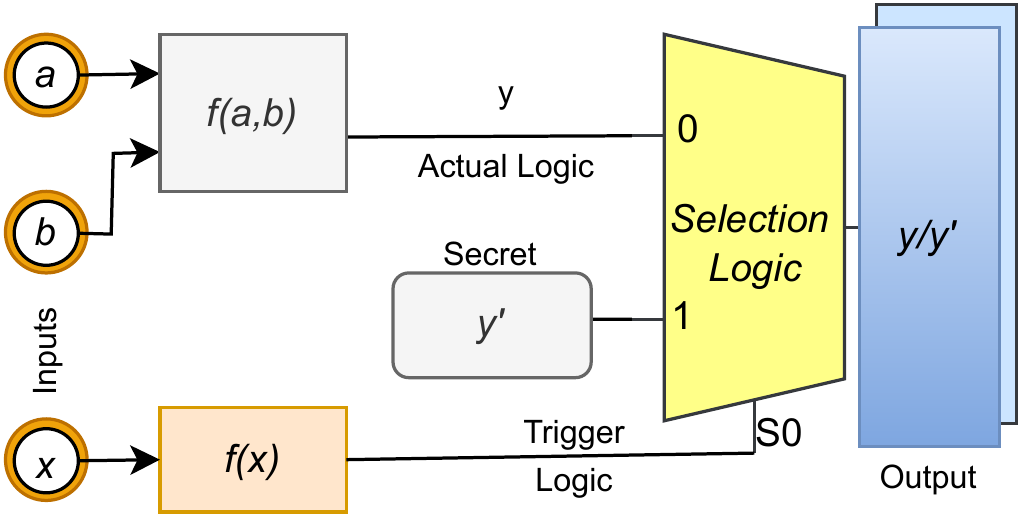}}
\caption{An implementation of a typical Trojan---a trigger will cause selection logic to overwrite the actual output logic with some secret modified output logic.}
\label{fig:trojanblock}
\end{figure}

{\flushleft \bf Trojans} have a broad existing taxonomy marked by several common factors. 
Firstly, all Trojans are modifications to a design in some way in the pursuit of some hidden goal. These will typically be stealthy modifications, and usually have the goal of either (1) compromising a target's functionality, or (2) adding additional functionality to leak or exfiltrate information / intellectual property. 
While it is more common for Trojans to exist in the software and information system domains, concern for their presence in other domains, particularly hardware-based, is growing.
Fig.~\ref{fig:trojanblock} presents a system diagram of a generic Trojan with a trigger and a payload. The trigger activates the Trojan logic when supplied with a particular combination of inputs \textit{a} and \textit{b}. In this example, the Trojan then overwrites the actual system logic with some secret Trojan logic.

In the software domain, traditional vectors for Trojans include malicious insiders and computer hackers. Still, there is also increasing attention being placed on Trojans that may be introduced via third-party developers in the ever-growing software supply chain
(for example, The SolarWinds hack~\cite{NYT_SolarWinds}, discovered in December 2020, was a significant supply-chain cyber attack that targeted various organizations, including government agencies and private companies). 
In embedded systems, built with first-party and third-party firmware (in this work, we use `firmware' to refer to embedded `software'; `firmware Trojans' are therefore a subset of `software Trojans'), the use of these kinds of supply chains is also growing.
For instance, many embedded devices are now being programmed with third-party and complex middleware and operating systems~\cite{borgohain_survey_2015} (e.g. mbed OS, FreeRTOS), increasing the difficulty for full source audits.

Meanwhile, in the hardware domain, widespread use of third-party IPs, usage of outsourced designs and test services, and vendor-provided Electronic Design Automation (EDA) tools all provide opportunities for adversaries to insert malicious modifications in a given design (Hardware Trojans~\cite{5406669,mcguire2019pcb}).
Given the increasing complexities of state of the art hardware, such Trojans are typically hard to detect, even in the presence of pre-silicon design verification~\cite{abramovici2009integrated}. Post-silicon verification involves de-packaging, and here too reverse-engineering to determine the presence of Trojans has challenges, including scalability~\cite{4484928}. 
A notable example of a Hardware Trojan was Bloomberg's reported `Big Hack'~\cite{robertson_big_2018} / `Long Hack'~\cite{robertson_long_2021}, which alleged that hardware Trojans were being introduced in the PCB supply chain~\cite{paley_active_2016, mcguire2019pcb, piliposyan2020hardware} by rogue manufacturers who were modifying computer motherboards. Though Bloomberg's claims were met with skepticism, such attacks have been demonstrated as possible~\cite{hudson_modchips_2018}, hence the research need.

{\flushleft \bf Triggers:} Typically, Trojans activate their malicious behaviours in the response to a `trigger'~\cite{chakraborty2009hardware}. In both hardware and software,
this trigger can come from internal or external sources based either on digital or analog origins (or a combination of both)~\cite{westhoff_concealed_2006}, such as temperature ~\cite{jang_secret_2015}, system inputs ~\cite{keuninckx_encryption_2017}, and positional data ~\cite{abouelnour2022situ}.
In general, Trojan activation is rare, which complicates detection.

{\flushleft \bf Payload:} The payload of a Trojan refers to the ultimate effects of the Trojan: for instance the ability to change or leak values~\cite{chakraborty2009hardware}. Examples in hardware include Trojans that alter the logic value of internal cells or memory registers; software Trojans may introduce backdoors or steal secrets. 
Trojans vary in size: software Trojan modifications can be as small as a single line of code or configuration, and simple hardware Trojans can have payloads as simple as a single gate, wire, or additional component to impact the delay or power consumption of the overall circuit. 
Trojans may also be used to abet side-channel attacks, particularly in the hardware domain where examples have been shown that leak information through power traces~\cite{4484928}, thermal radiation, and optical modulations of an output LED~\cite{defcon}. 
Cross-domain Trojans exist, whereby a hardware Trojan may result in changes to software (e.g. the aforementioned `Big Hack'~\cite{robertson_big_2018, robertson_long_2021, mehta2020big} which had a hardware component change embedded firmware). 
While software Trojans are broad in their capabilities, hardware Trojans can be classified in a more restricted sense (at least to the limit where they themselves might introduce software changes), where triggers and payloads can be separated into the digital and analog domains~\cite{chakraborty2009hardware}.

\subsection{DM and Cybersecurity}
Digital Manufacturing (DM) has become increasingly applicable (and commercially attractive) to various industries over the past decade. The primary benefit of DM involves simplifying manufacturing by decoupling product designers from product manufacturers~\cite{westhoff_concealed_2006, keuninckx_encryption_2017, wang_fast_2019} (i.e. Manufacturing-as-a-Service) and increasing the intelligence of manufacturing processes and production lines (`Smart manufacturing')~\cite{paritala_digital_2017}.
Such transformations will allow for small and medium-scale enterprises to federate; utilizing general purposes machines for producing a variety of product lines and workflows with complex shapes, textures, properties, and functionalities~\cite{mahesh2020survey}.
Here, digital production files will move between designers and manufacturers across networks, creating communication channels between all actors in a product's lifecycle: designers, OEMs, and vendors; as well as the individual machines that make up the product's production line.
This kind of network and `continuous streaming' of data enables intelligent data-driven decision-making for guiding modifications and calibration.

AM ~\cite{wong2012review} is a subset of DM where products are constructed layer-by-layer via techniques such as Fused Filament Fabrication (FFF), allowing for more complex part geometry and decreasing prototyping time. AM is also enabling manufacturing at the product's consumption point (on-site manufacturing) as well as product customization, e.g. for medicine~\cite{drugs} and prosthetics~\cite{heartvalves, bones} which may be specific to a particular customer's requirements ~\cite{rani2017digital, kuznetsov2019information, chenQR, yampolskiy2021did}.

As DM grows in complexity and connectivity, it becomes more difficult to secure it against all types of cyberattacks ~\cite{mahesh2020survey}. 
In particular, DM systems are cyber-physical systems (CPS), integrations of physical and cyber (digital) components networked together.
For instance, consider a FFF 3D printer: its physical components include embedded microprocessors running code, a heat bed and heaters for the filament, motors for moving the print head, sensors for detecting positional and temperature data, and support infrastructure for each of these categories (such as power supplies, etc.). Firmware code that runs in the microprocessors then makes up the cyber component---this would be responsible for receiving print instruction (probably as \texttt{g-code}) and then converting those instructions into commands for the physical components.
This firmware would be loaded using other networked cyber-devices, such as a general-purpose PC or remote server; the instructions for printing will also arrive in this manner.

The process of converting the design file into instructions for the printer will be completed digitally: products are first designed with Computer-aided design (CAD) software before being `sliced' for the 3D printer. Typical file formats will include models defined in \texttt{.stl} format as input to the `slicer'.
The slicer software is a Computer Aided Manufacturing (CAM) tool which breaks down 3D models into construction code---in the case of AM, these will be layers to print. Slicing software (and similar) will allow for specifying various parameters like infill rate, layer thickness, extrusion rate, fan speed, nozzle travel speed, etc. These parameters are crucial for correct manufacturing. 

Whereas traditional manufacturing plants were siloed with `air gaps' ensuring a relative level of security~\cite{tuptuk_security_2018}, it is immediately apparent that the DM approach will instead allow for various and new kinds of remote threat actors, especially given the trend of 
As such, comprehensive threat models must consider the potential vulnerabilities in the components making up DM and the interfaces between these components.

\begin{figure}[b]
\centering
\centerline{\includegraphics[scale=1]{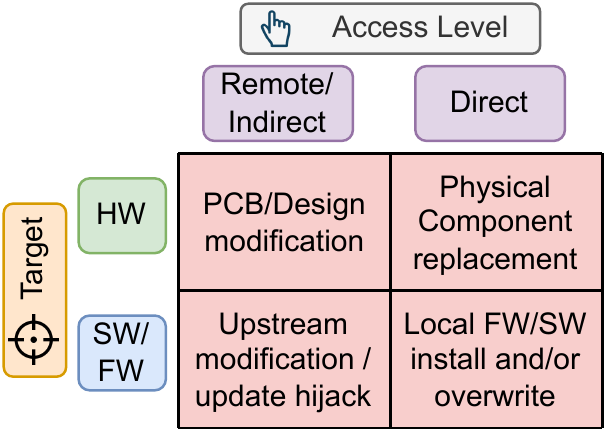}}
\caption{Mapping example Trojan attacks on Digital Manufacturing via Access/Target properties.}
\label{fig:pointofattack}
\end{figure}

\begin{figure*}[t]
  \includegraphics[scale=0.65]{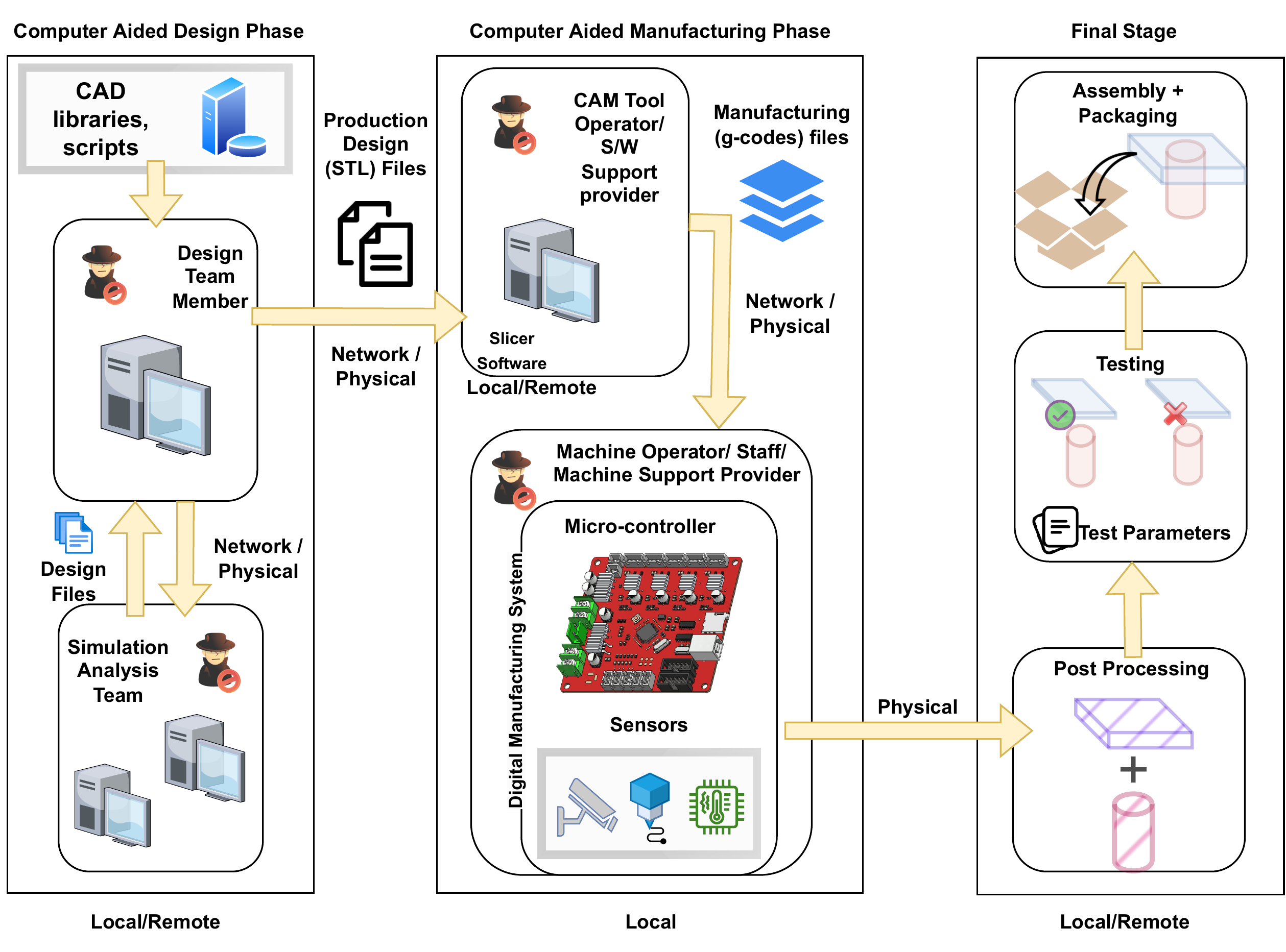}
  \caption{A generalized Digital Manufacturing process. Possible Hardware and software Trojans locations are indicated.}
  \rule{17cm}{0.6pt}
  \label{fig:AttackSurface}
\end{figure*}

\subsection{Overview of the DM Trojan attack surface}

\subsubsection{Direct and Indirect access}
Potential malicious changes to DM processes and equipment (Trojans) can be categorized according to the \textit{access level required} as well as the \textit{targets}, although combinations of these are possible.
We present an example of this in Fig.~\ref{fig:pointofattack}. 
In general, we can scope Trojan modifications as requiring either \textit{direct} or \textit{indirect} (remote) access to a given system.

Direct access would imply the adversary has real-world physical access to the plants and machinery they are targeting and allows for attacks requiring such access.
In the hardware domain, this could include direct modifications to a machine, including replacing or damaging components (e.g., power supplies, sensors, motors, and cameras). Even apparently minor notifications, such as to gears or nozzles, could have a significant impact on production quality. 
In the software/firmware domain, backdoors or deliberate software bugs/defects could be introduced by directly reinstalling firmware or software. This level of access might be necessary if remote updates are unavailable or machines are disconnected from any networks (`air gapped').
Trojan attacks may also be cross-domain: for instance, an adversary that inserts a physical device (e.g., a USB memory device) into an embedded PC (e.g., inside a DM machine), which then performs some software-based attack.

Meanwhile, indirect access covers the threats still possible by fully-remote adversaries (an adversary with full access can perform both local and remote-style attacks, but this is not true in reverse).
Here, hardware attacks will be limited to those in the supply chain: perhaps by remotely altering design files or OEM instructions. Still, these can result in large changes to the performance of a system.
In addition, software attacks are often not-so-limited, especially if a target system is capable of remote updates or reconfiguration. This is increasingly common for modern equipment but poses a major threat vector where adversaries can replace software or change calibration data.

\subsubsection{Attacks on DM supply chains}
The distributed nature of DM exposes production supply chains to threats, both cyber and physical.
In general, we divide these chains into three phases.
We present these phases, the agents within, and the communication channels between them in Fig.~\ref{fig:AttackSurface}.
The communication systems used between agents in these systems also provide avenues for interference.

The first is the Computer Aided Design (CAD) phase.
Here, a design is developed and simulated. This could be performed by a single person, or team, or divided amongst several specialized teams in single or multiple organizations. The utilized software may be generic (emails, instant messengers between colleagues) or specialized (CAD programs, finite element analysis simulators).
The eventual product will be design files for production, often in STL files.
Here, malicious modifications might change or steal product geometry (e.g., \cite{belikovetsky2017dr0wned}) or simulation results.

The second phase is Computer Aided Manufacturing (CAM), where the design files are converted into instruction files suitable for the various process equipment utilized---often in the form of \texttt{g-code}. Again, this task may be simple or may be complex, scaling from individual files and team members to large numbers of parts in various qualities and large organizations.
Once again, malicious agents may interfere with this, introducing subtle defects (e.g. \cite{beckwith2021needle}). Any form of manipulation of the \texttt{g-code}, like changing the amount of extrusion, changing the fan settings, changing the heating settings of the heatbed, or changing the X and Y coordinates, can severely impact the quality of the finished product.

The third phase, post-processing, has a limited threat space from hardware and software Trojans, although machine or process instructions may still face interference---especially those that deal with product testing and validation.

The most vulnerable threat locations exist on boundaries between phases and fundamental processes, as there may not be detailed information flows operating in the reverse direction.
For instance, after an STL production file is converted into \texttt{g-code}, it is not straightforward to obtain the original production file (the slicing process is \textit{lossy}).
These kinds of transitions thus allow for subtle malicious changes to be inserted. 
Further, these design and production files are also valuable for malicious actors seeking to steal IP, as they encapsulate high quality data about the products and production process. 
We explore these threats in more detail in Section~\ref{sec:threatmodel}.

\subsection{The MITRE ATT\&CK for ICS framework}

The MITRE ATT\&CK framework for Industrial Control Systems (ICS) is a comprehensive tool for understanding the tactics, techniques, and procedures used by adversaries in targeting industrial control systems. This framework, proposed by MITRE~\cite{alexander2020mitre}, is a collection of curated knowledge about various cyber-adversarial attacks on ICS that have occurred over decades. Their threat matrix categorizes threats based on the behavior of the cyber adversaries. It breaks down the attack into several stages, starting from the intrusion into the system to the final execution of the attack, and lists the different possible actions that an attacker is most likely to take in each of these stages. This allows organizations to better protect their industrial control systems by implementing appropriate countermeasures and defense strategies. Additionally, the framework can also be used to assess the effectiveness of current security measures and identify any gaps that need to be addressed.
Given DM is a subset of ICS, many of the concepts in the ATT\&CK framework are relevant, and could be utilized by DM manufacturing organizations.

\section{Trojan Threat Landscape for DM}
\label{sec:threatmodel}
\subsection{Overview}
As noted, the cyber-physical nature of the DM process exposes it to a variety of different threats across the production life-cycle; from third-party CAD tools and software to PCBs with potentially counterfeit and malicious on-board components.
An attacker can target any part of the DM system to achieve their malicious aims.

\begin{figure*}[h!]
\centering
\centerline{\includegraphics[scale=0.58]{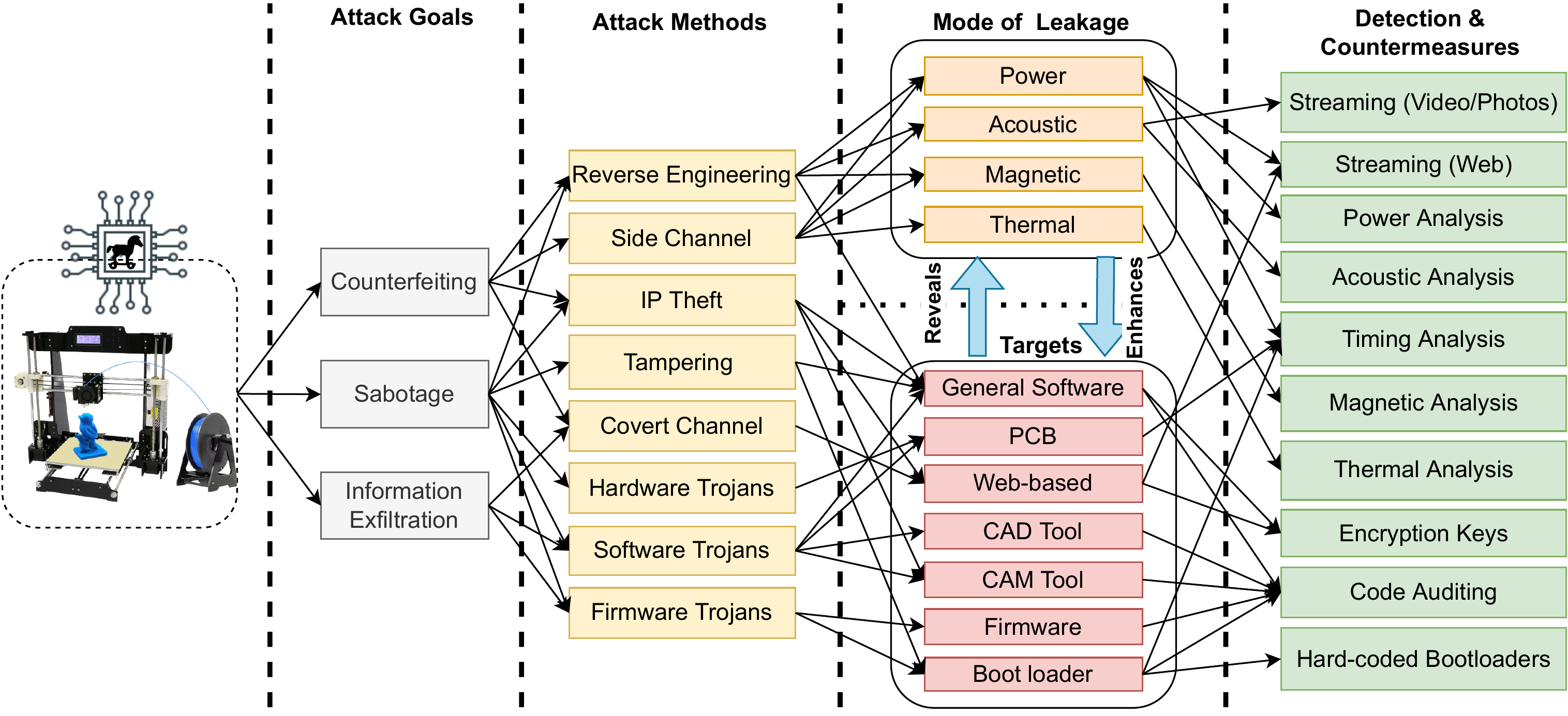}}
\caption{Threat taxonomy for Trojans impacting Additive Manufacturing.}
\label{fig:threatmodel}
\rule{17cm}{0.6pt}
\end{figure*}

Any defensive strategy will rely on constructing attack models and risk profiles for a given product and process. 
This requires the threat landscape to be well understood.
We thus categorise the known attack goals, attack methods, attack targets and the related modes of information leakage into a single threat landscape. 
In the rest of this section we will detail each element in this taxonomy.
Then, in Section~\ref{sec:detection} we will discuss detection and countermeasure strategies for these attacks.
The resultant model of this categorization is presented in Fig.~\ref{fig:threatmodel}.

\subsection{Attack goals} 

An attacker can have various goals when targeting a 3D printing system, which are typically motivated by financial or strategic gain.
It is important to note that the goals are not mutually exclusive, and an attacker may have multiple objectives when targeting a DM system. 
In general, when considering attacks on IP, the attackers are seeking out access to design and manufacturing files (e.g. STL and \texttt{g-code}).

We organize the attack goals into three broad classes:

\subsubsection{Intellectual Property Theft / Information Exfiltration}
Here, attackers are seeking to steal valuable IP. This may be in the form of product designs, production processes or calibration data, organizational data or other trade secrets.
Theft of this data could harm an organization's competitive edge, or in certain cases cause reputational harm if the information is made public / leaked. 
Common agents here include malicious insiders who seek to exfiltrate data, or remote adversaries who find weaknesses in a corporation's defensive firewalls.
While in the general case access to information can be protected by normal information security practices, hardware and software Trojans within DM can cause out-of-band information leakage; for instance in the form of side-channel leakage. 
This has been demonstrated using steganography to hide sensitive data in 3D models~\cite{rani2017digital, kuznetsov2019information, chenQR, yampolskiy2021did}.

\subsubsection{Counterfeiting} Although this can be considered a specific case of IP theft, counterfeiting has additional properties that make it a separate class of attacker goal.
Here, attackers are not necessarily interested in obtaining digital data on a given product; they instead wish to obtain additional fraudulent copies of that product. 
The major risk of counterfeits is their lack of provenance and (in general) the lack of quality control during their production: they are often sold without proper vetting of their design parameters (which may not be available to the counterfeiter; or may be too expensive or otherwise undesirable to perform).
As such, counterfeits are usually of inferior quality and pose safety risks, undermining the original product's integrity and causing harm to customers. 
An example of counterfeiting is demonstrated in~\cite{song2016my,al2016acoustic,al2016forensics}, where the authors use various side-channels to gather information about the object being printed. This information can later be used to manufacture poor quality and potentially risky counterfeits.

\subsubsection{Sabotage / Impacting Product Quality}
With the most insidious attack goal, saboteurs seek to introduce deliberate defects to production lines and products. They may lower print quality~\cite{pearce2022flaw3d} or modify design files to feature subtle mistakes \cite{belikovetsky2017dr0wned}.
Consequences range from weakened and defective parts to physical damage to manufacturing equipment. 
Denial of service attacks which block the printing process physically or remotely and  delay the production or irrecoverably damage  the product can be thought of as sabotage. ~\cite{beckwith2021needle} and~\cite{charalampous2021vision} demonstrated the importance of implementing detection methods to mitigate the effects of  attacks.

\begin{table*}
\centering
\begin{tabular}{|l|c|c|ccc|cccccccc|}
\hline
\multirow{2}{*}{\textbf{Papers}} & \multirow{2}{*}{\rotatebox{90}{Attacks~~~~~~~~~~}} & \multirow{2}{*}{\rotatebox{90}{Defenses~~~~~~~~~}} & \multicolumn{3}{c|}{\textbf{Attack Goals}} & \multicolumn{8}{c|}{\textbf{Attacks}} \\ \cline{4-14} 
 &  &  & \multicolumn{1}{l|}{\rotatebox{90}{Sabotage}} & \multicolumn{1}{l|}{\rotatebox{90}{Counterfeit~~~~}} & \begin{tabular}[c]{@{}l@{}} \rotatebox{90}{\shortstack[l]{Information\\exfiltration}}\end{tabular} & \multicolumn{1}{l|}{\begin{tabular}[c]{@{}l@{}}\rotatebox{90}{\shortstack[l]{Reverse\\Engineering}}\end{tabular}} & \multicolumn{1}{l|}{\begin{tabular}[c]{@{}l@{}}\rotatebox{90}{\shortstack[l]{Side -\\Channel}}\end{tabular}} & \multicolumn{1}{l|}{\begin{tabular}[c]{@{}l@{}}\rotatebox{90}{IP theft}\end{tabular}} & \multicolumn{1}{l|}{\begin{tabular}[c]{@{}l@{}} \rotatebox{90}{\shortstack[l]{Data\\Tampering}}\end{tabular}} & \multicolumn{1}{l|}{\begin{tabular}[c]{@{}l@{}} \rotatebox{90}{\shortstack[l]{Covert\\Channel}}\end{tabular}} & \multicolumn{1}{l|}{\begin{tabular}[c]{@{}l@{}} \rotatebox{90}{\shortstack[l]{Hardware\\Trojan}}\end{tabular}} & \multicolumn{1}{l|}{\begin{tabular}[c]{@{}l@{}} \rotatebox{90}{\shortstack[l]{Software\\Trojan}}\end{tabular}} & \begin{tabular}[c]{@{}l@{}} \rotatebox{90}{\shortstack[l]{Firmware\\Trojan}}\end{tabular} \\ \hline

\multicolumn{1}{|l|}{Pearce et. al.(FLAW3D) \cite{pearce2022flaw3d}} & \multicolumn{1}{c|}{\checkmark} & \multicolumn{1}{c|}{\checkmark} & \multicolumn{1}{c|}{\checkmark} & \multicolumn{1}{l|}{} & \multicolumn{1}{l|}{} & \multicolumn{1}{l|}{} & \multicolumn{1}{l|}{} & \multicolumn{1}{l|}{} & \multicolumn{1}{c|}{\checkmark} & \multicolumn{1}{l|}{} & \multicolumn{1}{l|}{} & \multicolumn{1}{l|}{} & \checkmark \\ \hline

\multicolumn{1}{|l|}{Moore et. al. (Desktop printer S/W) \cite{moore2016vulnerability}} & \multicolumn{1}{l|}{} & \multicolumn{1}{c|}{\checkmark} & \multicolumn{1}{l|}{} & \multicolumn{1}{l|}{} & \multicolumn{1}{c|}{\checkmark} & \multicolumn{1}{c|}{\checkmark} & \multicolumn{1}{l|}{} & \multicolumn{1}{c|}{\checkmark} & \multicolumn{1}{c|}{\checkmark} & \multicolumn{1}{l|}{} & \multicolumn{1}{l|}{} & \multicolumn{1}{c|}{\checkmark} &  \\ \hline

\multicolumn{1}{|l|}{Caleb et. al. (Haystack) \cite{beckwith2021needle}} & \multicolumn{1}{l|}{} & \multicolumn{1}{c|}{\checkmark}  & \multicolumn{1}{c|}{\checkmark} & \multicolumn{1}{l|}{} & \multicolumn{1}{l|}{} & \multicolumn{1}{l|}{} & \multicolumn{1}{l|}{} & \multicolumn{1}{l|}{} & \multicolumn{1}{c|}{\checkmark} & \multicolumn{1}{l|}{} & \multicolumn{1}{l|}{} & \multicolumn{1}{l|}{} & \checkmark \\ \hline

\multicolumn{1}{|l|}{Song et. al. (Smartphone) \cite{song2016my}} & \multicolumn{1}{c|}{\checkmark} & \multicolumn{1}{l|}{} & \multicolumn{1}{l|}{} & \multicolumn{1}{c|}{\checkmark} & \multicolumn{1}{l|}{} & \multicolumn{1}{c|}{\checkmark} & \multicolumn{1}{c|}{\checkmark} & \multicolumn{1}{l|}{} & \multicolumn{1}{l|}{} & \multicolumn{1}{l|}{} & \multicolumn{1}{l|}{} & \multicolumn{1}{l|}{} &  \\ \hline

\multicolumn{1}{|l|}{Gatlin et. al. (Detect power signature) \cite{gatlin2019detecting}} & \multicolumn{1}{l|}{} & \multicolumn{1}{c|}{\checkmark} & \multicolumn{1}{c|}{\checkmark} & \multicolumn{1}{l|}{} & \multicolumn{1}{l|}{} & \multicolumn{1}{l|}{} & \multicolumn{1}{c|}{\checkmark} & \multicolumn{1}{l|}{} & \multicolumn{1}{l|}{} & \multicolumn{1}{l|}{} & \multicolumn{1}{l|}{} & \multicolumn{1}{l|}{} &  \\ \hline

\multicolumn{1}{|l|}{Gao et. al. (ThermoTag) \cite{gao2021thermotag}} & \multicolumn{1}{l|}{} & \multicolumn{1}{c|}{\checkmark} & \multicolumn{1}{l|}{} & \multicolumn{1}{c|}{\checkmark} & \multicolumn{1}{l|}{} & \multicolumn{1}{l|}{} & \multicolumn{1}{l|}{} & \multicolumn{1}{c|}{\checkmark} & \multicolumn{1}{l|}{} & \multicolumn{1}{l|}{} & \multicolumn{1}{l|}{} & \multicolumn{1}{l|}{} &  \\ \hline

\multicolumn{1}{|l|}{Gao et. al.(Watch and safeguard) \cite{gao2018watching}} & \multicolumn{1}{l|}{} & \multicolumn{1}{c|}{\checkmark} & \multicolumn{1}{c|}{\checkmark} & \multicolumn{1}{l|}{} & \multicolumn{1}{l|}{} & \multicolumn{1}{l|}{} & \multicolumn{1}{l|}{} & \multicolumn{1}{l|}{} & \multicolumn{1}{l|}{} & \multicolumn{1}{l|}{} & \multicolumn{1}{l|}{} & \multicolumn{1}{l|}{} & \checkmark \\ \hline

\multicolumn{1}{|l|}{Mohammad et. al. (Acoustic SCA) \cite{al2016acoustic}} & \multicolumn{1}{c|}{\checkmark} & \multicolumn{1}{l|}{} & \multicolumn{1}{l|}{} & \multicolumn{1}{c|}{\checkmark} & \multicolumn{1}{l|}{} & \multicolumn{1}{c|}{\checkmark} & \multicolumn{1}{l|}{} & \multicolumn{1}{c|}{\checkmark} & \multicolumn{1}{l|}{} & \multicolumn{1}{l|}{} & \multicolumn{1}{l|}{} & \multicolumn{1}{l|}{} &  \\ \hline

\multicolumn{1}{|l|}{Mahesh et. al. (Survey) \cite{mahesh2020survey}} & \multicolumn{1}{c|}{\checkmark} & \multicolumn{1}{c|}{\checkmark} & \multicolumn{1}{c|}{\checkmark} & \multicolumn{1}{c|}{\checkmark} & \multicolumn{1}{l|}{} & \multicolumn{1}{c|}{\checkmark} & \multicolumn{1}{c|}{\checkmark} & \multicolumn{1}{c|}{\checkmark} & \multicolumn{1}{c|}{\checkmark} & \multicolumn{1}{c|}{\checkmark} & \multicolumn{1}{l|}{} & \multicolumn{1}{l|}{} & \checkmark \\ \hline

\multicolumn{1}{|l|}{Faezi et. al (Forensics Thermal SC) \cite{al2016forensics}} & \multicolumn{1}{c|}{\checkmark} & \multicolumn{1}{l|}{} & \multicolumn{1}{l|}{} & \multicolumn{1}{c|}{\checkmark} & \multicolumn{1}{l|}{} & \multicolumn{1}{l|}{} & \multicolumn{1}{c|}{\checkmark} & \multicolumn{1}{c|}{\checkmark} & \multicolumn{1}{l|}{} & \multicolumn{1}{l|}{} & \multicolumn{1}{l|}{} & \multicolumn{1}{l|}{} &  \\ \hline

\multicolumn{1}{|l|}{Gupta et. al. (Supply chain CS) \cite{gupta2020additive}} & \multicolumn{1}{l|}{} & \multicolumn{1}{c|}{\checkmark} & \multicolumn{1}{c|}{\checkmark} & \multicolumn{1}{c|}{\checkmark} & \multicolumn{1}{l|}{} & \multicolumn{1}{c|}{\checkmark} & \multicolumn{1}{c|}{\checkmark} & \multicolumn{1}{l|}{} & \multicolumn{1}{c|}{\checkmark} & \multicolumn{1}{l|}{} & \multicolumn{1}{l|}{} & \multicolumn{1}{l|}{} &  \\ \hline

\multicolumn{1}{|l|}{Belikovetsky et. al. (dr0wned) \cite{belikovetsky2017dr0wned}} & \multicolumn{1}{c|}{\checkmark} & \multicolumn{1}{l|}{} & \multicolumn{1}{c|}{\checkmark} & \multicolumn{1}{l|}{} & \multicolumn{1}{l|}{} & \multicolumn{1}{l|}{} & \multicolumn{1}{l|}{} & \multicolumn{1}{l|}{} & \multicolumn{1}{c|}{\checkmark} & \multicolumn{1}{l|}{} & \multicolumn{1}{l|}{} & \multicolumn{1}{l|}{} &  \\ \hline

\multicolumn{1}{|l|}{Yampolskiy et. al. (3Dprinter weapon) \cite{yampolskiy2016using}} & \multicolumn{1}{c|}{\checkmark} & \multicolumn{1}{l|}{} & \multicolumn{1}{c|}{\checkmark} & \multicolumn{1}{l|}{} & \multicolumn{1}{l|}{} & \multicolumn{1}{l|}{} & \multicolumn{1}{l|}{} & \multicolumn{1}{l|}{} & \multicolumn{1}{l|}{} & \multicolumn{1}{l|}{} & \multicolumn{1}{l|}{} & \multicolumn{1}{l|}{} &  \\ \hline

\multicolumn{1}{|l|}{Yampolskiy et. al. (Attack taxonomy) \cite{yampolskiy2018security}} & \multicolumn{1}{c|}{\checkmark} & \multicolumn{1}{c|}{\checkmark} & \multicolumn{1}{c|}{\checkmark} & \multicolumn{1}{c|}{\checkmark} & \multicolumn{1}{l|}{} & \multicolumn{1}{l|}{} & \multicolumn{1}{l|}{} & \multicolumn{1}{l|}{} & \multicolumn{1}{l|}{} & \multicolumn{1}{l|}{} & \multicolumn{1}{l|}{} & \multicolumn{1}{l|}{} &  \\ \hline

\multicolumn{1}{|l|}{Yampolskiy et. al. (Steganography) \cite{yampolskiy2021did}} & \multicolumn{1}{c|}{\checkmark} & \multicolumn{1}{l|}{} & \multicolumn{1}{l|}{} & \multicolumn{1}{l|}{} & \multicolumn{1}{c|}{\checkmark} & \multicolumn{1}{l|}{} & \multicolumn{1}{l|}{} & \multicolumn{1}{l|}{} & \multicolumn{1}{l|}{} & \multicolumn{1}{c|}{\checkmark} & \multicolumn{1}{l|}{} & \multicolumn{1}{c|}{\checkmark} &  \\ \hline

\multicolumn{1}{|l|}{Keuninckx et. al. (Chaos sync.) ~\cite{keuninckx_encryption_2017}} & \multicolumn{1}{l|}{} & \multicolumn{1}{c|}{\checkmark} & \multicolumn{1}{c|}{\checkmark} & \multicolumn{1}{l|}{} & \multicolumn{1}{c|}{\checkmark} & \multicolumn{1}{l|}{} & \multicolumn{1}{l|}{} & \multicolumn{1}{l|}{} & \multicolumn{1}{l|}{} & \multicolumn{1}{l|}{} & \multicolumn{1}{l|}{} & \multicolumn{1}{l|}{} &  \\ \hline

\multicolumn{1}{|l|}{Wang et. al. (Fast encryption) ~\cite{wang_fast_2019}} & \multicolumn{1}{l|}{} & \multicolumn{1}{c|}{\checkmark} & \multicolumn{1}{c|}{\checkmark} & \multicolumn{1}{l|}{} & \multicolumn{1}{c|}{\checkmark} & \multicolumn{1}{c|}{\checkmark} & \multicolumn{1}{l|}{} & \multicolumn{1}{l|}{} & \multicolumn{1}{l|}{} & \multicolumn{1}{l|}{} & \multicolumn{1}{l|}{} & \multicolumn{1}{l|}{} &  \\ \hline

\multicolumn{1}{|l|}{Durumeric et. al. (HTTPS certificate) ~\cite{durumeric_analysis_2013}} & \multicolumn{1}{l|}{} & \multicolumn{1}{c|}{\checkmark} & \multicolumn{1}{c|}{\checkmark} & \multicolumn{1}{c|}{\checkmark} & \multicolumn{1}{c|}{\checkmark} & \multicolumn{1}{l|}{} & \multicolumn{1}{l|}{} & \multicolumn{1}{l|}{} & \multicolumn{1}{c|}{\checkmark} & \multicolumn{1}{l|}{} & \multicolumn{1}{l|}{} & \multicolumn{1}{l|}{} &  \\ \hline

\multicolumn{1}{|l|}{Westhoff et. al. Data aggregation) ~\cite{westhoff_concealed_2006}} & \multicolumn{1}{l|}{} & \multicolumn{1}{c|}{\checkmark} & \multicolumn{1}{l|}{} & \multicolumn{1}{c|}{\checkmark} & \multicolumn{1}{c|}{\checkmark} & \multicolumn{1}{l|}{} & \multicolumn{1}{l|}{} & \multicolumn{1}{l|}{} & \multicolumn{1}{l|}{} & \multicolumn{1}{l|}{} & \multicolumn{1}{l|}{} & \multicolumn{1}{l|}{} &  \\ \hline

\multicolumn{1}{|l|}{Jang et. al. (SeCRet) ~\cite{jang_secret_2015}} & \multicolumn{1}{l|}{} & \multicolumn{1}{c|}{\checkmark} & \multicolumn{1}{l|}{} & \multicolumn{1}{l|}{} & \multicolumn{1}{c|}{\checkmark} & \multicolumn{1}{c|}{\checkmark} & \multicolumn{1}{l|}{} & \multicolumn{1}{l|}{} & \multicolumn{1}{l|}{} & \multicolumn{1}{c|}{\checkmark} & \multicolumn{1}{l|}{} & \multicolumn{1}{l|}{} &  \\ \hline

\multicolumn{1}{|l|}{Kanter et. al. (Syncing of Random Bit) ~\cite{kanter_synchronization_2010}} & \multicolumn{1}{l|}{} & \multicolumn{1}{c|}{\checkmark} & \multicolumn{1}{c|}{\checkmark} & \multicolumn{1}{l|}{} & \multicolumn{1}{c|}{\checkmark} & \multicolumn{1}{l|}{} & \multicolumn{1}{l|}{} & \multicolumn{1}{l|}{} & \multicolumn{1}{l|}{} & \multicolumn{1}{l|}{} & \multicolumn{1}{l|}{} & \multicolumn{1}{l|}{} &  \\ \hline

\multicolumn{1}{|l|}{Chai et. al. (Image encryption) ~\cite{chai_novel_2017}} & \multicolumn{1}{l|}{} & \multicolumn{1}{c|}{\checkmark} & \multicolumn{1}{l|}{} & \multicolumn{1}{l|}{} & \multicolumn{1}{l|}{} & \multicolumn{1}{c|}{\checkmark} & \multicolumn{1}{l|}{} & \multicolumn{1}{l|}{} & \multicolumn{1}{l|}{} & \multicolumn{1}{l|}{} & \multicolumn{1}{l|}{} & \multicolumn{1}{l|}{} &  \\ \hline

\multicolumn{1}{|l|}{Tonello et. al. (Secret key exchange) ~\cite{tonello_secret_2015}} & \multicolumn{1}{l|}{} & \multicolumn{1}{c|}{\checkmark} & \multicolumn{1}{c|}{\checkmark} & \multicolumn{1}{l|}{} & \multicolumn{1}{l|}{} & \multicolumn{1}{l|}{} & \multicolumn{1}{l|}{} & \multicolumn{1}{l|}{} & \multicolumn{1}{l|}{} & \multicolumn{1}{l|}{} & \multicolumn{1}{l|}{} & \multicolumn{1}{l|}{} &  \\ \hline

\multicolumn{1}{|l|}{Goh et. al. (Review on ML methods) ~\cite{Goh}} & \multicolumn{1}{l|}{} & \multicolumn{1}{c|}{\checkmark} & \multicolumn{1}{c|}{\checkmark} & \multicolumn{1}{l|}{} & \multicolumn{1}{l|}{} & \multicolumn{1}{l|}{} & \multicolumn{1}{c|}{\checkmark} & \multicolumn{1}{l|}{} & \multicolumn{1}{c|}{\checkmark} & \multicolumn{1}{l|}{} & \multicolumn{1}{l|}{} & \multicolumn{1}{l|}{} &  \\ \hline

\multicolumn{1}{|l|}{Holzmond et. al. (In-situ detection) ~\cite{HOLZMOND2017135}} & \multicolumn{1}{l|}{} & \multicolumn{1}{c|}{\checkmark} & \multicolumn{1}{c|}{\checkmark} & \multicolumn{1}{l|}{} & \multicolumn{1}{l|}{} & \multicolumn{1}{l|}{} & \multicolumn{1}{l|}{} & \multicolumn{1}{l|}{} & \multicolumn{1}{c|}{\checkmark} & \multicolumn{1}{l|}{} & \multicolumn{1}{l|}{} & \multicolumn{1}{l|}{} &  \\ \hline

\multicolumn{1}{|l|}{Khan et. al. (real-time detection)~\cite{FARHANKHAN2021521}} & \multicolumn{1}{l|}{} & \multicolumn{1}{c|}{\checkmark} & \multicolumn{1}{c|}{\checkmark} & \multicolumn{1}{l|}{} & \multicolumn{1}{l|}{} & \multicolumn{1}{l|}{} & \multicolumn{1}{l|}{} & \multicolumn{1}{l|}{} & \multicolumn{1}{c|}{\checkmark} & \multicolumn{1}{l|}{} & \multicolumn{1}{l|}{} & \multicolumn{1}{l|}{} &  \\ \hline

\multicolumn{1}{|l|}{Wu et. al. (Detecting Malicious Defects)~\cite{wu2016detecting}} & \multicolumn{1}{l|}{} & \multicolumn{1}{c|}{\checkmark} & \multicolumn{1}{c|}{\checkmark} & \multicolumn{1}{l|}{} & \multicolumn{1}{l|}{} & \multicolumn{1}{l|}{} & \multicolumn{1}{l|}{} & \multicolumn{1}{l|}{} & \multicolumn{1}{c|}{\checkmark} & \multicolumn{1}{l|}{} & \multicolumn{1}{l|}{} & \multicolumn{1}{l|}{} &  \\ \hline

\multicolumn{1}{|l|}{Wu et. al. (Detecting CPA)~\cite{Wu}} & \multicolumn{1}{l|}{} & \multicolumn{1}{c|}{\checkmark} & \multicolumn{1}{c|}{\checkmark} & \multicolumn{1}{l|}{} & \multicolumn{1}{l|}{} & \multicolumn{1}{l|}{} & \multicolumn{1}{c|}{\checkmark} & \multicolumn{1}{l|}{} & \multicolumn{1}{c|}{\checkmark} & \multicolumn{1}{l|}{} & \multicolumn{1}{l|}{} & \multicolumn{1}{l|}{} &  \\ \hline

\multicolumn{1}{|l|}{Chen et. al. (Anomaly Detection)~\cite{Chen}} & \multicolumn{1}{l|}{} & \multicolumn{1}{c|}{\checkmark} & \multicolumn{1}{c|}{\checkmark} & \multicolumn{1}{l|}{} & \multicolumn{1}{l|}{} & \multicolumn{1}{l|}{} & \multicolumn{1}{l|}{} & \multicolumn{1}{l|}{} & \multicolumn{1}{l|}{} & \multicolumn{1}{l|}{} & \multicolumn{1}{l|}{} & \multicolumn{1}{l|}{} &  \\ \hline

\multicolumn{1}{|l|}{Moore et. al. (Firmware Implication) ~\cite{moore_implications_2017}} & \multicolumn{1}{c|}{\checkmark} & \multicolumn{1}{l|}{} & \multicolumn{1}{c|}{\checkmark} & \multicolumn{1}{l|}{} & \multicolumn{1}{l|}{} & \multicolumn{1}{l|}{} & \multicolumn{1}{l|}{} & \multicolumn{1}{l|}{} & \multicolumn{1}{l|}{} & \multicolumn{1}{l|}{} & \multicolumn{1}{l|}{} & \multicolumn{1}{l|}{} & \checkmark \\ \hline

\multicolumn{1}{|l|}{Pearce et. al. (PCB Trojans \& detection) ~\cite{pearce2022detecting}} & \multicolumn{1}{c|}{\checkmark} & \multicolumn{1}{c|}{\checkmark} & \multicolumn{1}{l|}{\checkmark} & \multicolumn{1}{c|}{} & \multicolumn{1}{l|}{} & \multicolumn{1}{l|}{} & \multicolumn{1}{c|}{\checkmark} & \multicolumn{1}{l|}{} & \multicolumn{1}{l|}{} & \multicolumn{1}{l|}{} &  \multicolumn{1}{c|}{\checkmark} & \multicolumn{1}{c|}{\checkmark}  &  \\ \hline

\end{tabular}
\caption{Classification of state-of-the-art DM Trojan attacks and defenses based on attack goals and methods.
\label{tab:papers}}
\rule{17cm}{0.6pt}
\end{table*}

\subsection{Attack Methods} An attacker can employ multiple methods to achieve their goals: 

\subsubsection{Reverse Engineering} In reverse engineering, the attacker seeks to obtain design information from observation of artifacts of the supply chain. This can involve using the finished product, or may consist of re-constructing inputs to design stages given the outputs of later phases.
    
\subsubsection{Side-channel attacks} Here, attackers are seeking to utilize the natural side-channel emissions from a DM system to elucidate secrets about the DM product under production.
These attacks range from expensive acoustic side-channel~\cite{al2016acoustic} and thermal side-channel~\cite{al2016forensics} attacks to comparatively inexpensive and practical smartphone-based attacks~\cite{song2016my}.

\subsubsection{IP theft} Once information is discovered or made available, stealing it may require its own steps, especially if some kind of exfiltration process is required. Moore et. al.~\cite{moore2016vulnerability} explores several software vulnerabilities that can lead to IP theft.
            
\subsubsection{Tampering} This involves the malicious alterations of files or data loaded, stored, received, or sent during the manufacturing process, primarily utilized to sabotage the printing operation. Prominent examples include changes to \texttt{g-code}~\cite{beckwith2021needle} and STL files~\cite{belikovetsky2017dr0wned}.

\subsubsection{Covert Channel} The establishment of one or more covert channels allows for the undetected transfer of secrets outside of a DM process. An example using stegonography was demonstrated where STL files were utilized as secret channels~\cite{yampolskiy2021did}.
    
\subsubsection{Hardware, Software, and Firmware Trojans}
The methods explored by this survey; Trojans introduced in the software~\cite{moore2016vulnerability}, firmware~\cite{moore_implications_2017,pearce2022flaw3d} and hardware~\cite{pearce2022detecting} have demonstrated the ability to leak information, sabotage products and compromise equipment.

\subsection{Attack Targets} 
Attack Targets define where the manifestations of an attack will appear. These are closely interrelated with DM side chanels, so Fig.~\ref{fig:threatmodel} presents them together---an attacked component will have its side channels change in potentially detectable manners. Certain attacks/Trojans will leverage this, using the side channels as a channel for information leakage.
As with goals and methods, an individual attack may feature more than one target from the following:

\subsubsection{General software} 
General-purpose computer equipment and software is used extensively in most DM systems. Attacks may therefore target commodity applications and tools such as Microsoft Windows, Linux, and so on. 

\subsubsection{PCB/Hardware}
The electronic hardware that makes up DM production system lines is a potential target for hardware Trojans, especially those that may modify production data or enhance side-channel leakage~\cite{pearce2022detecting}.

\subsubsection{Web-based software}
Given the propensity of DM production files to travel over computer networks and the web, web-based software for file transfers and communication is a target that could yield considerable IP and opportunities for file manipulation.

\subsubsection{CAD Tool/software}
When designing digital products, designers typically utilize domain-specific and third-party software to generate and simulate the design (for instance using finite element analysis, fluid dynamic analysis or multi-physics analysis). If these programs are compromised, attackers may gain total control over the target IP. Even small defects introduced to the software could compromise the ability for the target to reliably generate design and production data.
However, modifications here may have to circumvent traditional information security methods such as anti-malware and anti-virus scans.

\subsubsection{CAM Tool/software}
This category, which includes proprietary machine-operating software as well as tools for file preparation (e.g. slicing software), is responsible for DM processes that prepare DM produts for manufacturing.
Infections here could modify parts or production orders~\cite{moore2016vulnerability,yampolskiy2021did}.
If this software is contained within an embedded system, standard information security practices may not be as prevalent (for instance, an embedded Windows or Linux instance running without anti-malware protections).

\subsubsection{Firmware}
Firmware runs on the embedded controllers of DM machines such as AM printers. The firmware is responsible for the proper execution of command instructions (e.g. \texttt{g-code}), and is responsible for sending and receives signals to control the various physical components of the machine, such as motors, heaters, and sensors.
A compromised firmware~\cite{moore_implications_2017} can modify commands in-flight to compromise the printing process (and even potentially damage the hardware). 

\subsubsection{Bootloader}
Separately to the firmware are lower-level programs responsible for installation and debugging. It has also been demonstrated that these are a valid target for attack~\cite{pearce2022flaw3d}. Whereas normal firmware may feature updates, bootloaders will rarely need to change over the lifecycle of a machine ~\cite{kamhoua2014testing}, and may also avoid standard audit procedures. This means that a bootloader compromised as part of a supply chain attack could remain undetected.

\subsection{Side-channel leakage attacks}

\subsubsection{Acoustic Side Channel:}
DM involves the movement of various actuators and motors, which results in the emission of distinct and audible signatures. If captured using a microphone, these can provide information useful for reverse engineering.
Demonstrated examples include on AM, where motion in the x,y, and z axis were classified using ML~\cite{song2016my}; and ML-based shape extraction with low-quality microphones~\cite{al2016acoustic}. Printers may even be able to be uniquely identified by their audio signatures~\cite{hojjati2016leave}.

\subsubsection{Magnetic Side Channel}
DM systems such as 3D printers may also be leak magnetic information from stepper motors. When each motor in a system rotates at a different rate, they will produce distinct magnetic signatures obtainable via a magnetometer; which may be able to elucidate information such as directions and speed~\cite{song2016my,bilal2017review,chhetri2019tool}.

\subsubsection{Thermal Side Channel}
Thermal side-channel attacks on DM systems utilize the different thermal signatures emitted by `hot' components (such as the nozzle, heat bed, and extruded material in AM) to gather information about the manufacturing process. When captured by thermal cameras, these signatures can reveal sensitive information such as the shape and dimensions of the printed object, the material used, and the printing parameters~\cite{bilal2017review,islam2017exploiting}. Careful analysis can even reverse-engineer production \texttt{g-code} and potentially steal IP~\cite{al2016forensics,tsalis2019taxonomy}. 

\subsubsection{Visual Side Channel:}
Many DM systems such as high-end 3D printers are now provided with pre-installed cameras used to monitor the manufacturing process, allowing remote examining of any intrusions or hindrances. Although undoubtably useful, the presence of a camera capturing each step of the manufacturing process of an object can be a major source of information leakage. Based on the camera feeds from a compromised camera, the design may be reconstructed, leading to potential IP theft. 

\subsubsection{Power Side Channel:}
DM systems typically utilize considerable power for their various components and actuators; microprocessors, motors, and heaters amongst them.
Non-invasive inductive current measuring techniques can thus be used to capture side-channel data and may result in leakage of information about the manufacturing process---even potentially leaking encryption keys where data has been properly protected~\cite{islam_pmu-trojan_2018}.

\subsection{Key takeaways}
Using the Trojan threat taxonomy from Fig.~\ref{fig:threatmodel}, and our discussion of each known state-of-the-art attack presented in this section, we can organize the literature into Table~\ref{tab:papers}.
We also note that the literature includes defenses for specific attacks and methods in digital manufacturing.
These defenses will be discussed further in Section~\ref{sec:detection}.

\section{Detection and Countermeasures}
\label{sec:detection}

In this section we will discuss the presented methods for detecting malicious modifications to DM processes. Table \ref{tab:countermeasures} presents published defensive mechanisms and the modes of attack they seek to protect against; including firmware and web-based systems.
As seen in the table, a common approach among the proposed countermeasures is using encryption keys to secure sensitive information. Moreover, some papers propose the use of code verification techniques to ensure that only authorized code is executed in DM systems. %

\begin{table}[t]
\centering
\footnotesize
\begin{tabular}{|l|l|l|l|} 
\toprule
\textbf{{Defense}}                                                        & \textbf{{Mode of attack}}                                   & \textbf{{Target}}                                                & \textbf{{Countermeasures}}                                                                                           \\ 
\toprule
Chaos sync. ~\cite{keuninckx_encryption_2017}                                                                    &Web-based                                                           & Firmware                                                                & Encryption Keys                                                                                                             \\ 
\hline
Fast encryption ~\cite{wang_fast_2019}                                                                & Web-based                                                           & Firmware                                                                & \begin{tabular}[c]{@{}l@{}}Encryption Keys, \\ Code Verification\end{tabular}                                               \\ 
\hline
HTTPS certificate ~\cite{durumeric_analysis_2013}                                                              & Web-based                                                          & \begin{tabular}[c]{@{}l@{}}Software\\/Firmware\end{tabular}                                                               & Encryption Keys                                                                                                             \\ 
\hline
Data aggregation ~\cite{westhoff_concealed_2006}                                                               & Web-based                                                          & Firmware                                                                & Encryption Keys                                                                                                             \\ 
\hline
TEEs/SeCRet ~\cite{jang_secret_2015}                                                                         &  Web-based                                                         & Firmware                                                                & \begin{tabular}[c]{@{}l@{}}Encryption Keys, \\ Code Verification\end{tabular}                                               \\ 
\hline
\begin{tabular}[c]{@{}l@{}}Vision based ~\cite{LIES201829, HOLZMOND2017135, charalampous2021vision} \\error detection  \end{tabular}         & \begin{tabular}[c]{@{}l@{}}g-code \\ manipulation\end{tabular}     & \begin{tabular}[c]{@{}l@{}}Software\\/Firmware\\ /Hardware\end{tabular} & Camera Feeds                                                                                                                \\ 
\hline
A review on ML~\cite{Goh}                                                                   & \begin{tabular}[c]{@{}l@{}}g-code\\ manipulation\end{tabular}      & \begin{tabular}[c]{@{}l@{}}Software\\/Firmware\\ /Hardware\end{tabular} & \begin{tabular}[c]{@{}l@{}}Camera Feeds,\\ Machine Learning\end{tabular}                                                    \\ 
\hline
\begin{tabular}[c]{@{}l@{}}Inline 3D print \\failure~\cite{Lyngby}\end{tabular}                & \begin{tabular}[c]{@{}l@{}}g-code\\ manipulation\end{tabular}      & \begin{tabular}[c]{@{}l@{}}Software\\/Firmware\\ /Hardware\end{tabular} & \begin{tabular}[c]{@{}l@{}}Camera Feeds, \\ Reference Models\end{tabular}                                                   \\ 
\hline
Audio signatures~\cite{belikovetsky2019digital} & \begin{tabular}[c]{@{}l@{}}Acoustic \\ side channel\end{tabular}   & \begin{tabular}[c]{@{}l@{}}Software\\/Firmware\\ /Hardware\end{tabular} & \begin{tabular}[c]{@{}l@{}}Comparison of \\audio fingerprints\end{tabular}                                                   \\ 
\hline
KCAD ~\cite{kcadacousticdet}                                                                           & \begin{tabular}[c]{@{}l@{}}Acoustic \\ side channel\end{tabular}   & \begin{tabular}[c]{@{}l@{}}Software\\/Firmware\\ /Hardware\end{tabular} & \begin{tabular}[c]{@{}l@{}}Additional moves \\while printing\end{tabular}                                                   \\ 
\hline
\begin{tabular}[c]{@{}l@{}}Detecting ~\cite{gatlin2019detecting}\\Sabotage ~\end{tabular}                   & \begin{tabular}[c]{@{}l@{}}g-code\\ manipulation\end{tabular}      & \begin{tabular}[c]{@{}l@{}}Software\\/Firmware\\ /Hardware\end{tabular} & \begin{tabular}[c]{@{}l@{}}Compares with \\golden Power \\Signature, \\Machine learning\end{tabular}                        \\ 
\hline
Thermotag ~\cite{gao2021thermotag}                                                                      &    \begin{tabular}[c]{@{}l@{}}g-code\\ manipulation\end{tabular}                                                                &   \begin{tabular}[c]{@{}l@{}}Software\\/Firmware\\ /Hardware\end{tabular}                                                                      & \begin{tabular}[c]{@{}l@{}}Leverages uniqueness\\ of the 3D printer by \\Fingerprinting\end{tabular}                        \\ 
\hline
Secure PCB ~\cite{bhunia2014securepcb}                                                                     & hardware trojan                                                    & Hardware                                                                & X-ray                                                                                                                       \\ 
\hline
Big Hack prevention~\cite{mehta2020big}                                                                   & hardware trojan                                                    & Hardware                                                                & \begin{tabular}[c]{@{}l@{}}Big data analysis \\on PCB qualities\end{tabular}                                                \\ 
\hline
\begin{tabular}[c]{@{}l@{}}Software ~\cite{moore2016vulnerability}\\vulnerabilities\\ detection \end{tabular} & \begin{tabular}[c]{@{}l@{}}software/\\firmware trojan\end{tabular} & \begin{tabular}[c]{@{}l@{}}Software\\/Firmware\end{tabular}             & \begin{tabular}[c]{@{}l@{}}code analysis, \\code auditing,\\ code updation,\\ restricted sharing, \\debugging\end{tabular}  \\ 
\hline
Watermarking ~\cite{rani2017digital}                                                                   & Software trojan                                                    & Software                                                                & \begin{tabular}[c]{@{}l@{}}Bar code, \\QR code\end{tabular}                                                                 \\
\bottomrule
\end{tabular}
\caption{Published defensive methods suitable for DM}
\label{tab:countermeasures}
\vspace{-8mm}
\end{table}

\subsection{Web-based countermeasures}
\subsubsection{HTTPS/TLS} Utilzing the secure counterpart of hypertext transfer protocol (HTTP), HTTPS, is a standard information security best-practice. Here, communication is protected by the Transport Layer Security (TLS) protocol, which uses an asymmetric public key framework with private and public keys for data encryption~\cite{durumeric_analysis_2013}. 
TLS is often already seen in well-made networked embedded devices to protect their communication. %

\subsection{Encryption-based countermeasures}
\subsubsection{Chaos Synchronization} 
Deriving encryption keys for protecting data can be non-trivial. One proposal uses a transmitter and receiver connected to a `chaotic driver' via a public channel which is responsible for generating such keys and ensuring their relative uniqueness~\cite{keuninckx_encryption_2017, kanter_synchronization_2010, tonello_secret_2015}. In DM, it may thus be possible to use chaotic synchronization of encryption keys to limit access between a device and the printer. 

\subsubsection{Fast Encryption}: Wang et al.~\cite{wang_fast_2019, chai_novel_2017} discuss using chaotic synchronization to deploy encryption keys and code verification. Here, models are converted into two-dimensional objects for matrix encryption. Unlike a chaotic driver, this method uses a lossless lattice-mapping format. Similar to the method of chaotic synchronization, fast encryption could be used to further scramble the original keys used to protect DM files.  

\subsubsection{Concealed Data Aggregation in Wireless Sensor Networks (WSNs)} In their research, Westhoff et al. in~\cite{westhoff_concealed_2006} focus on achieving end-to-end encryption for Concealed Data Aggregation in Wireless Sensor Networks (WSNs). The technique involves transforming the encrypted data before it passes through the system. The information is separated, strengthened, and recombined when it reaches its destination. %
This type of data management and encryption would be helpful in large-scale DM, where large networks of devices work together with potentially-sensitive production and customer data.%

\subsubsection{TEEs/SeCRet} 
Trusted Execution Environments (TEE) can be used to ensure the protection of security-sensitive resources from malicious code~\cite{sabt_trusted_2015}; and could prevent attacks such as the bootloader FLAW3D attack~\cite{pearce2022flaw3d}.
However, in isolation, TEE may not be enough to protect communication  between legitimate software processes.
In DM, where many different kinds of process will be occurring and requiring the secure transfer of data, it may be possible to utilize the SeCReT secure communication channels defined by Jang et al.~\cite{jang_secret_2015}.
Here, channels for communication of the data utilize unique session keys for authentication and integrity purposes. These keys are unique and cannot be retrieved by an attacker, as they are immediately expelled from the system once they have been utilized.

\subsection{Camera feed based countermeasures}
With the diversity of existing and emerging types of Trojans, developing new detection methods for each attack is inefficient. Due to improving camera technology (higher resolutions and frame rates), many vision-based side-channel detection methods have been proposed for DM---particularly in AM, where cameras can straightforwardly capture the layer-by-layer construction of printed parts~\cite{refId0}. 
Studies have thus developed vision-based detection methods to monitor 3D printer components such as extruder condition~\cite{LIES201829}, as well as monitor the printed parts~\cite{HOLZMOND2017135, charalampous2021vision}. Most studies concentrate on monitoring the printed object itself, as it is the most prominent and direct way to identify mistakes or anomalous results. The anomalies that can be detected using vision-based methods include missing or extra layers~\cite{pr8111464}, incorrect layer thickness~\cite{CARRASCOCORREA2021116177}, and misaligned or distorted shapes~\cite{Kadam}. 
Anomaly detection in digital manufacturing typically involves several different methods, including machine learning algorithms that analyze data from cameras~\cite{FARHANKHAN2021521, ROSSI2021438, gohil2022deterrent}, a reference model to compare the current printed object~\cite{Charalamous, Petsiuk}, and statistical analysis to identify unusual patterns or deviations from expected results~\cite{He} (this may occur when production files have been deliberately altered, possibly by Trojans).
Here we discuss the different types of vision-based anomaly detection methods:

\subsubsection{Machine Learning} 
Machine Learning methods for detecting defects are prominent in recent studies. Models can learn from a large amount of data and detect different Trojans with good accuracy ~\cite{Wu, gohil2022deterrent}. These methods may be relatively fast and even suitable for in-situ additive manufacturing monitoring ~\cite{Goh}.
However, there are also some disadvantages of machine learning methods. 
First, collecting that extensive data needed for training a model can be expensive (in terms of both time and materials). Finished models may lack flexibility, only detecting defects that they were trained to identify. Given the variability in process manufacturing, there is also the risk of false positives~\cite{wu2016detecting} where standard manufacturing/printing variations are flagged as defects and disrupt the build process.

\subsubsection{Reference Models}
Reference models are typically used to identify defects in manufacturing and build processes. They can help identify deviations or abnormal results potentially caused by malicious embedded Trojans. In vision-based detection methods, the reference model serves as a digital representation of the expected result, which includes details such as its properties, shape, and dimensions. Within AM, this reference model method divides into two major categories: layer-wise detection~\cite{Lyngby} and completed product detection~\cite{Zhao}. Nowadays, studies on defect detection in AM tend to focus more on layer-wise inspection, as this approach allows a more comprehensive comparison between the reference model and the printed object. Rather than just looking at the exterior, the layer-wise examination can provide a more detailed analysis of the printed structure and properties.

\subsubsection{Statistical Analysis}
Statistical Analysis techniques are often utilized in vision-based anomaly detection. This method typically involves calculating and analyzing statistical measures of designated areas in the photos. The Python OpenCV~\cite{OpenCV} library is often used for image processing~\cite{Chen}. Statistical methods can help detect abnormal results in large datasets. However, it is sensitive to the choice of measured features and threshold values to determine whether results are anomalous.

\subsection{Other side-channel based countermeasures}\label{subsec:sidechanneldetection}

\subsubsection{Audio} Just as the sound emissions caused by digital manufacturing processes can be leveraged by attackers to reveal sensitive information, it may also be used to detect anomalous behavior caused by Trojans. Audio fingerprints may be captured and analyzed for differences between prints~\cite{belikovetsky2019digital}.
Alternatively, to obfuscate the audio leakage, additional (redundant) moves may be made in the 3D space (e.g. \cite{kcadacousticdet} counters the audio attack from \cite{al2016acoustic}).

\subsubsection{Power Analysis} 

Power side-channel analysis has been widely used for detecting hardware Trojans. Two different sources of power leakage can be used for such analysis:
{\flushleft \it PCBs:}  In \cite{piliposyan2020hardware}, the authors suggest recording the dynamic and static power consumption of a golden model of the PCB board. The presence of a Trojan in a PCB under test will be confirmed if there is a significant deviation in the current PCB's dynamic and static power profile from the golden PCB.

 {\flushleft \it Motors:}  In \cite{gatlin2019detecting}, the authors suggest recording the power signatures of a particular printing operation performed by a golden model of the 3D printer. Later, these signatures are compared to the signature of the 3D printer under test, to determine any manipulation in the \texttt{g-code}, thereby detecting the presence of any firmware or hardware Trojans. 

\subsubsection{PCB Trojan detection using X-ray} PCB X-ray imaging is a powerful tool for identifying potential security issues in printed circuit boards. It allows for non-destructive examination of the internal layers of a PCB to reveal hidden faults or defects, such as missing or misaligned components, fused or disconnected traces, traces having inconsistent cross-sections, defects that may not be visible to the naked eye ~\cite{wang2022security}. These defects can cause device malfunction or something less obvious like early ageing of the circuit. In~\cite{bhunia2014securepcb,mcguire2019pcb}, the authors demonstrate the use of X-ray imaging for identifying the above potential security issues in PCBs and propose a method for automatically analyzing such issues. PCB X-ray is a valuable tool for identifying potential security issues in digital manufacturing and should be considered as part of an overall security strategy.

\subsubsection{PCB qualities/properties} The electrical properties of a Printed Circuit Board (PCB) is determined by several factors, including its resistance, capacitance, and other electrical characteristics. These factors play a critical role in determining the performance and functionality of a PCB. In~\cite{mehta2020big}, the authors propose a big data analytics approach to analyze PCB qualities, which can be used to predict the PCB's performance and identify potential quality issues. This approach can be helpful for organizations that need to quickly and efficiently assess the quality of their PCBs.

\subsubsection{Timing and other side-channels} These and other side-channels (such as CPU frequency/load/DVFS~\cite{islam_pmu-trojan_2018} and Hardware Performance Counters~\cite{basu_theoretical_2020}) may be used by Trojans or Trojan detection strategies. 
Given the cyclical nature of CPS, it is often possible to derive system arrangements such that measuring the side-channels via `loop-backs' will provide for Trojan discovery: for instance if a Trojan capable of editing outputs introduces delays when feeding forward even unaltered values~\cite{pearce2022detecting}.
In the DM case, it should be possible to construct these kinds of loop-backs to enable such timing measurements (one example could be a computer which visually monitors the print head position and only issues the next g-code once it reaches a set point).

\section{Taxonomy related case studies}
\label{sec:case-studies}
In this section, we will further explore the various attack vectors and countermeasures presented in Section~\ref{sec:threatmodel} and Section~\ref{sec:detection} via a series of six DM-based case studies.
These highlight the risks of Trojans in DM as well as provide insights into the state-of-the-art security solutions for this domain.

\subsection{Case Study 1: Information Exfiltration through Steganography} %
\subsubsection{Problem:} In a business setting, proprietary design IPs, classified documents, and files are considered valuable secrets of great interest to rival firms. These firms may collude with an insider within the targeted firm to exfiltrate these pieces of information. Insiders often try to leak information over the internet as most devices within a firm are directly or indirectly connected to it. To prevent such leakage, techniques such as internet traffic monitoring that looks out for suspicious message packets are often in place. As a result, insiders try to find innovative ways of communicating with the outside world without getting detected.

One such method is using a software Trojan that hides in plain sight and sends information to the outsider. Regular code monitoring and updates can make the insertion of such Trojans difficult. Another method that insiders may use is steganography, which tries to hide confidential information in some form inside the actual 3D model itself. In~\cite{rani2017digital}, the authors demonstrated information hiding by embedding 3D barcodes on a 3D object. In~\cite{kuznetsov2019information}, the author embeds information in 3D matrices inside a container 3D model. Separate colors are used in printing the embedded design and the container design such that it is not noticeable from the outside but distinguishable under a laser scan.

The embedded objects, though invisible from the outside, can reduce the quality and functional value of the printed object. In~\cite{chenQR}, the authors embedded components of the QR code in different design layers that are retrievable by a CT scan. In~\cite{yampolskiy2021did}, the authors demonstrate a technique of exfiltrating information without changing the aesthetics or quality of the printed object. This technique is based on manipulating the STL file, which breaks down the 3D design into a collection of facets which are triangles defined by the position of the vertices, and a normal that defines the direction. By varying the ordering of the vertices in each facet while maintaining the right-hand thumb rule constraint, the authors could encode information in the STL file without impacting the geometry of the printed object.
While~\cite{rani2017digital,chenQR} can be used as a physical water-marking techniques to validate the source of the file, and it can similarly be used as a channel to exfiltrate information. \cite{yampolskiy2021did} also used a software Trojan that uses STL files as a covert channel to leak secrets to outsiders.   

\subsubsection{Solution:} One solution to prevent steganography-based attacks like ~\cite{kuznetsov2019information} is to implement strict access controls on the STL files and other production files used in the 3D printing process. This can be done by implementing a secure file sharing system that only allows authorized personnel to access these files, or potentially encrypting information in storage/in transit.  Another solution involves implementing a 3D scanning system to detect any embedded objects or patterns within the printed object. Such 3D scans can be implemented using techniques such as CT scan or laser scan that helps in detecting any printing anomalies or hidden object embedded within the actual object. Comparing the power signature with the golden model can also reveal possible modifications in the STL or \texttt{g-code} ~\cite{gatlin2019detecting}.

\subsection{Case Study 2: Firmware Trojan Sabotage} %
\subsubsection{Problem:} The \texttt{g-code} generated by CAM software tools such as Cura 3D~\cite{cura}, ReplicatorG ~\cite{replicatorg}, and Repetier-Host~\cite{repetier-host} are sent to 3D printers through various methods, including USB drives, card readers, and network connections. The firmware running on the controller parses this \texttt{g-code} to extract information such as layer thickness, fan mode, extrusion length, X, Y, and Z axis coordinates, heatbed temperature, and more. These instructions are crucial for the successful printing of the desired end product. However, their importance also makes them a target for attackers.

Previous research has analyzed the potential vulnerabilities in 3D printer software and firmware, such as the Marlin firmware ~\cite{Marlin}, which was found to have weak coding practices such as using \texttt{strcat} with fixed global buffers that can cause a buffer overflow and unsafe use of signed/unsigned comparison along with \texttt{strlen}, which can lead to integer overflow ~\cite{moore2016vulnerability}. In ~\cite{moore_implications_2017}, the authors demonstrated two attacks on 3D printing firmware. The first attack involves loading malicious Marlin firmware to the 3D printer controller, which hardcodes instructions to print a pyramid instead of the intended cube. The second attack modifies a private variable in the Marlin code that controls the extruder feed rate, resulting in the printing of structurally deformed cube structures.

Another concern is the firmware's bootloader, responsible for uploading, downloading, and verifying the firmware. Despite users frequently reloading the firmware, the bootloader is rarely updated. This makes the bootloader an ideal target for Trojan insertion, as observed in \cite{pearce2022flaw3d}. The authors successfully injected a Trojan into the flash memory where the bootloader resides, using the bootloader Interrupt Service Routine(ISR) to change the value of the IVSEL register and free up memory space by altering the values of the SP and SH stack pointers. The malicious code then makes changes to the incoming \texttt{g-code} stored in a ring buffer, resulting in two attacks: reducing the amount of material extruded in a linear movement of the extruder and identifying sequences of G1(extrusion) commands and turning a subset of them into G0(no extrusion) commands, which significantly reduces the quality of the printed materials.

\subsubsection{Solution:} The code-related vulnerabilities of ~\cite{moore2016vulnerability} can be easily mitigated using secure coding practices. The malicious modification of the firmware code in ~\cite{moore_implications_2017} can be detected in an off-chip debugger. Attacks like printing alternate models, over-extrusion of ~\cite{moore_implications_2017}, and material reduction of ~\cite{pearce2022flaw3d} can be easily identified visually. A secure bootloader can implement data integrity and cryptographic checks to prevent such malicious changes in the firmware. The relocated material attack mentioned in \cite{pearce2022flaw3d} is comparatively hard to detect. The Trojan, if inserted successfully, covers its tracks by changing the variable values to the initial state before verification. Using an off-chip debugger to detect irregular ISR accesses and side-channel analysis are the probable solutions to this threat. Introduced delays in the ISRs may also be able to be detected in a loop-back timing setting~\cite{pearce2022detecting}. %

\subsection{Case Study 3: Hardware Trojan in PCBs} 
\subsubsection{Problem:} A hardware Trojan can be a significant threat to the security of 3D printers and additive manufacturing systems. As described in ~\cite{bhunia2014securepcb, wang2022security}, a hardware Trojan can be injected inside the PCB of a 3D printer by modifying the Printed Circuit Boards(PCB) or adding an extra component at the manufacturing time. These hardware Trojans can be used to steal sensitive information or disrupt the printer's operation, which can have serious consequences for the security and integrity of the additive manufacturing process.

\subsubsection{Solution:} To mitigate the risk of hardware Trojan attacks, it is essential to develop methods for detecting them if present and preventing the introduction of these malicious components in the first place. Insertion of a hardware Trojan often manifests as an increase in power, timing, or thermal side channels. The detection technique proposed in \cite{pearce2022detecting} uses a combination of static and dynamic analysis of various side channels like timing, power and magnetic side-channels to identify hardware Trojans in an embedded PCB. A similar approach can be extended to detect and prevent the introduction of hardware Trojans in the 3D printer PCB, which can help to ensure the security and integrity of the digital manufacturing process.

\subsection{Case Study 4: Trojan Targeting PCB Printers}
\subsubsection{Problem:} Printed Circuit Board (PCB) printers, such as the BotFactory SV2 ~\cite{botfactory}, have become a popular DM tool in both research and industry settings. These printers use AM to construct PCBs by using conducting and insulating inks in a layer-by-layer fashion.
This allows PCBs to be assembled by end-users at point-of-demand. Using PCB printers  opens the door to novel potential security threats.
For instance, a Trojan in these printers may be able to `replicate', by modifying printed electronics to include Trojans with their own malicious functionality! %

\subsubsection{Solution:} Similar to Case Study 2 and 3, Trojan attacks on PCB printers can primarily be avoided via secure firmware and software updates using techniques such as secure boot and code signing to ensure that only authorized updates are installed on the printer. A hardware-based security mechanism, such as a secure TEE, can also be implemented to store encryption keys and other sensitive information. %
Given the target (printed PCBs), scanning for PCB trojans in the production outputs would likely also be pertinent (e.g. using X-rays~\cite{bhunia2014securepcb}).

\subsection{Case Study 5: Reverse Engineering via\\Thermal Side-channel Leakage} %
\subsubsection{Problem:} In a study by Al et al. ~\cite{al2016forensics}, the potential for exploiting thermal signatures emitted by a 3D printer to reverse-engineer the \texttt{g-code} of an object being printed was demonstrated. The authors utilized a Printrbot 3D printer~\cite{PrintrBot} and a thermal camera placed 0.95 meters away from the printer. The camera was positioned in the $yz$ and $xz$ planes to capture the top-view and side-view of the printer. By tracking the movements of the nozzle-end and base plate via video feeds, the authors were able to map the movements to their corresponding \texttt{g-code} commands.

The authors used thermal imaging to detect the thermal signatures emitted by a 3D printer while in operation. They could reverse-engineer the \texttt{g-code} used to print the object by capturing the thermal signature. This technique was shown to be effective for different types of 3D printers, including the Printrbot used in the study. Additionally, the authors demonstrated the ability to reconstruct \texttt{g-code} for more complex shapes, such as a pentagon, which is a crucial step towards the potential use of this technique in forensic investigations~\cite{al2016forensics}.

\subsubsection{Solution:} The authors proposed two methods for this task: the static approach involving precise distance calculations and the dynamic approach utilizing a trained learning model. The dynamic method was more robust and less prone to errors as it can adjust for slight variations in the thermal signatures~\cite{al2016forensics}. They also proposed using this technique in forensic investigations to reconstruct the \texttt{g-code} from the thermal signatures, which can help identifying the attacker who tampered with the 3D printing process. The authors also suggested using this technique in various other applications like manufacturing and supply chains to secure 3D printing.

\subsection{Case Study 6: \texttt{g-code} modifications \\without a reference model}
\subsubsection{Problem:} In ~\cite{beckwith2021needle}, the authors used the blue-team/red-team approach, where the red team tampered with the printing instructions, and the blue team attempted to detect the changes. After converting CAD models to \texttt{g-code} with the slicer program, edits were made to the printing \texttt{g-code} to simulate subtle changes caused by Trojans, aiming to introduce defects in production files. Identifying modified files is not straightforward without access to the original model data or ground truth on what the correct \texttt{g-code} should be.

\subsubsection{Solution:} After noting common statistics about the \texttt{g-code} files, the blue team utilized statistical analysis and machine learning-based detection strategies for anomaly detection. Statistical analysis extracted features such as the number of G0 and G1 commands and the boundaries of X, Y, and Z values from \texttt{g-code} files. In machine-learning methods, Principal Component Analysis (PCA) and clustering algorithms were used to reduce dimensions and identify outliers.   

The results of the detection strategies employed by the blue team were promising, as the methods could detect 5 of the 6 types of defect introduced by the red team. However, the broader generalizability could still be investigated (applying the solutions to new kinds of Trojans). %

\subsection{Case Study 7: Camera Feed Analysis for\\Trojan detection} %
\subsubsection{Problem:} The possibility of the existence of various software and hardware Trojans in the 3D printing process requires the implementation of proper detection methods. But these methods are mostly used after the printing process has been fully completed, wasting time and printing materials. Without knowing the types of defects, it is hard to identify whether the printed result is tampered with. These defects are often camouflaged in the middle layers, and escape immediate detection. 

\subsubsection{Solution:}
In ~\cite{beckwith2021needle}, detection methods were utilized prior to transmitting the modified \texttt{g-code} to the 3D printer. However, the validity of the final printed object could not be confirmed. Through the implementation of vision-based methods, as outlined in ~\cite{charalampous2021vision} and also in ~\cite{abouelnour2022situ} it became feasible to detect a greater number of defects by comparing each printed layer with a reference model, thus enabling the detection of potentially malicious attacks on the printing process. With the variety of trojans emerging, detection methods with the flexibility to adapt to different situations can be helpful.

\section{Conclusions}
\label{sec:conclusions}

Digital Manufacturing (DM) represents a fundamental paradigm shift within the manufacturing sector: a shift away from siloed and air-gapped manufacturing systems highly-coupled between designers and OEMs, instead towards `smart' Manufacturing-as-a-Service environments where manufacturers will possess networks of general-purpose manufacturing processes and systems with the flexibility for many kinds of product geometry and capabilities. 
However, as presented in this work, the new generation of DM tools and processes does not come risk-free.
Hardware and software Trojans may enter the supply chain and design lifecycle of both products and production lines. They may leak sensitive and valuable data and sabotage products and equipment.

However, in order to build cohesive and comprehensive defensive strategies, attacker capabilities must be formalized and known. This paper thus surveyed the current state-of-the-art in this area, focusing on the demonstrated attacks and defenses in this domain.
We emphasize the need for implementing secure design and manufacturing practices: though manufacturing and cybersecurity are typically considered from two disparate viewpoints; their marriage is sorely needed in the DM space.

Security best-pratices exist in the traditional information security realm. Best-practices should also be developed for DM. Research and development should continue in this area, with the ultimate goal of protecting sensitive information, such as digital designs, from unauthorized access, exfiltration, and modification. Protections for production lines should likewise be comprehensive. Side-channel analysis and monitoring for anomalous behaviour should be standard, as well as careful protection of this data to protect against side-channel leakage. %
Future research could thus focus on this area: the opposing requirements for side-channel analysis without side-channel leakage will need novel techniques, especially if these techniques are to avoid disrupting or compromising the products under manufacture. %

Industries based in the DM realm should carefully consider the research presented in this survey. They are exposed to a vast quantity of threats, many novel, and many outside the normal scope of information security. 
Threat awareness in an organization can be achieved by increasing collaboration and communication between individuals and teams and by providing training and education about such threats. In addition, a full-fledged end-to-end monitoring system like the MITRE ATT\&CK for ICS would be suitable for handling this vast array of risks and determining a proper course of action to mitigate them.

\section*{Acknowledgments}

This research was supported in part by NSF Grant \#1931724. 
Any opinions, findings, and conclusions, or recommendations expressed are those of the author(s) and do not necessarily reflect the views of the National Science Foundation.

\bibliographystyle{ACM-Reference-Format}
\bibliography{references}

\end{document}